\documentstyle[epsf]{l-aa}

\begin{document}

\thesaurus{08(08.16.7)} 
   
\title{A Search for the Optical Counterpart of PSR B1951+32 in the Supernova Remnant CTB 80
\thanks{Based on observations taken at SAO, Zelenchuk, Karachai-Cherkessia, Russia}}


\author{Cr\'{e}idhe O'Sullivan,\inst{1} A. Shearer,\inst{2} M. Colhoun,\inst{2} A. Golden,\inst{1} M. Redfern,\inst{1} 
          R. Butler,\inst{3} G.M. Beskin,\inst{4} S.I. Neizvestny,\inst{4} V.V. Neustroev,\inst{4} V.L. Plokhotnichenko,\inst{4} 
          \& A. Danks, \inst{5}}

   \offprints{C O'Sullivan, creidhe@physics.ucg.ie}

   \institute{Department of Physics, National University of Ireland Galway, Ireland
        \and Information Technology Centre, National University of Ireland Galway, Ireland
        \and Department of Mathematics and Statistics, University of Edinburgh, Scotland
        \and SAO, Nizhnij Arhyz, Karachai-Cherkessia, Russia
        \and STX/Goddard Space Flight Centre, Greenbank, Maryland, US            }

   \date{Received ...; Accepted ...}

   \maketitle
   \markboth{O'Sullivan C. et al.: A Search for the Optical Counterpart of PSR B1951+32}{}


\begin{abstract}

Using time-resolved two-dimensional aperture photometry we have put upper limits
on the pulsed emission from two proposed optical counterparts for PSR
B1951+32.  Our pulsed upper limits of $m_{v,\rm{pulsed}} > 23.3$,
$m_{b,\rm{pulsed}} > 24.4$, for the first candidate and $m_{v,\rm{pulsed}} >
23.6$, $m_{b,\rm{pulsed}} > 24.3$ for the second, make it
unlikely that either of these is, in fact, the pulsar. We discuss three further
candidates, but also reject these on the basis of timing results. A search of a 
$5\farcs5 \times 5\farcs5$ area centred close to these stars failed to find any
 significant
pulsations at the reported pulsar period.

\keywords{  -- {\bf pulsars: individual:} PSR B1951+32
                
              }
\end{abstract}

\section{Introduction}

PSR B1951+32 is a  fast pulsar in the peculiar combination supernova
remnant (SNR) CTB 80.  The characteristic spin-down age of the pulsar
$(P/2\dot P)$ and the dynamical age of the SNR (Koo et al.  1990) are both
$\sim10^5$ yrs.  Its low surface magnetic field strength, of $~5\times10^7$ T, is
unusual amongst the selection of young and middle-aged pulsars detected so far but it has been
suggested that this is more representative of young pulsars in general
(\cite{lyne83}).  It is clearly important to test pulsar models with as wide
a variety of pulsar parameters as possible.

Steady emission from the pulsar has so far been detected at radio (\cite{strom87};
 \cite{kul88}) and X--ray (\cite{beck82}; \cite{wang84}) wavelengths.  Two possible
 optical counterparts have been proposed by Blair \& Schild (1985) and Fesen \&
 Gull (1985).  Since the initial discovery of the 39.5-ms pulsar (\cite{kul88}), evidence for pulsed emission has also been found in X--rays
 (\cite{safi95}) and $\gamma$--rays (Ramanamurthy et al.  1995) with upper
 limits in the infrared (Clifton et al. 1988).  To-date no upper limits on
 pulsed emission in the optical have been published.

Optical observations are essential in determining the relative contributions of 
thermal and magnetospheric processes to pulsar emission. The continuing improvements in 
optical-detector sensitivities means that pulsed emission from 6 isolated neutron
 stars has already been detected. In this paper we present time-resolved optical observations of the central region
 of the supernova remnant CTB 80 and, in particular, of the two proposed pulsar
 candidates.

\section{Observations}

The observations were made during 1996 June and 1997 June using NUI, Galway's TRIFFID
camera.  The TRIFFID system consists of a multianode microchannel array (MAMA)
2-dimensional photon-counting detector with a B extended S-20 photocathode
(\cite{tim85}) and a fast data-collection system (\cite{red93}).  The position
and time-of-arrival of each photon are recorded to a precision of 25$\mu$m and
1$\mu$s, respectively.  TRIFFID's 1024$\times$256-pixel array was mounted at the
prime focus of the SAO 6-m Telescope (BTA), resulting in an equivalent spatial
resolution of $0\farcs23$ pixel$^{-1}$.  Timing was achieved using a GPS
receiver and an ovened 10-MHz crystal.  Absolute timing was unavailable for the
1996 observations however, and times were recorded relative to the last reset of
the data-collection system.  The data-collection system was reset at the start
of the observing run and again after the 1996 June 23 observations.

The observations were made with standard B and V filters and are summarised in
Table 1.  In all, a total of 19,456 seconds of B-band data and 8,173 seconds of
V-band data were collected.  Photometric standard stars were observed each night
for calibration purposes.

\begin{table}
\caption[]{Summary of Observations}
\label{obs}
\begin{flushleft}
\begin{tabular}{ccccc}
\hline\noalign{\smallskip}
 Date          &  UTC    & Filter & Duration &  Seeing\\
               & \null   &  \null & (s)      & ($\arcsec$)       \\
\noalign{\smallskip}
\hline\noalign{\smallskip}
1996 June 22   & 20:31:00 &  B & 9013   & $1.40$ \\
1996 June 23   & 21:23:00 &  V & 3756   & $1.16$ \\
1996 June 23   & 22:28:30 &  B & 1333   & $1.28$ \\
1996 June 23   & 22:52:40 &  B & 3110   & $1.45$ \\
1996 June 24   & 22:33:41 &  V & 4417   & $1.40$ \\
1996 June 25   & 23:00:21 &  B & 3300   & $1.24$ \\
1997 June 04   & 22:59:17 &  B & 2700   & $1.50$ \\
\noalign{\smallskip}
\hline
\end{tabular}
\end{flushleft}
\end{table}

\section{Data Reduction}

The data were first binned into 1-ms frames and divided by a deep flatfield image
taken during the observation period.  A post-exposure shift-and-add sharpening
technique was applied to produce an integrated image (\cite{shear96}).  Because
of the large telescope aperture no significant improvement in the image above
the seeing limit resulted, but any artefacts due to effects such as telescope
wobble were removed.  Data from individual nights were summed to produced the
integrated V and B images shown in Figs. 1 and 2.

Photometry and astrometry were carried out using the IRAF daophot and GASP
packages, respectively.  The background level was taken to be the mean of
the signal in an annulus of radius $2\farcs0$ and width $0\farcs5$ centred on the
object position.  Stellar co-ordinates were calculated using the astrometry of
Blair \& Schild (1985).

Several pulsar candidate positions were chosen as described in Sec. 4.  Photon
times were extracted from a window with a diameter equal to the seeing width
centred on the pulsar candidate.  The time series was translated to the
solar-system barycentre using the JPL DE200 ephemeris and then folded in phase
using the PSR B1951+32 ephemeris of \cite{fost94} (Table 2).  The resulting
light curves were analysed using the $Z_{n}^{2}$ statistic (\cite{buc89}).  The
Fourier power-spectrum of each candidate was also calculated in the vicinity of
the first four harmonics.

\begin{table}
\caption[]{PSR B1951+32 Ephemeris (\cite{fost94})}
\begin{flushleft}
\begin{tabular}{ll}
\hline\noalign{\smallskip}
Parameter & Value\\
\noalign{\smallskip}
\hline\noalign{\smallskip}
$\nu$        &   25.29739673Hz \\
$\dot{\nu}$  &  $-3.73571\times10^{-12}$ Hz s$^{-1}$ \\
$\ddot{\nu}$ &  $-3.37\times10^{-23}$ Hz s$^{-2}$ \\
Epoch        &  2447005.6880\\
\noalign{\smallskip}
\hline
\end{tabular}
\end{flushleft}
\end{table}

\section{Results}

Figs. 1 and 2 show the integrated V- and B-band images.  The proposed optical
counterparts (\cite{blair85}; \cite{fes85}) labelled $\lq$1' and $\lq$2', can clearly be
seen in both images.  The core region surrounding the two candidates is shown in
Fig.  3, which is a sum of all the observations.  The positions of the radio
pulsar (\cite{fost94}) and X-ray point source (\cite{wang84}) are also marked.  Like Fesen \& Gull we find no evidence for Blair
and Schild's candidate $\lq$3' (down to a limiting magnitude of $m_v\approx24$), consistent
with its identification as a H$\alpha$+[NII] emission knot.  The photometry for
these candidates, and those discussed later on, is
given in Table 3.  The magnitudes were calibrated with respect to Blair and 
Schild's star $\lq$A'. The quoted errors were calculated from a combination of counting
statistics and flat-fielding errors.  Our B1950 source positions of the two
pulsar candidates are $\alpha=19^{h}51^{m}02\fs58,
\delta=+32^{\circ}44\arcmin50\farcs9$ (candidate 1) and $\alpha=19^{h}51^{m}02\fs49,
\delta=+32^{\circ}44\arcmin52\farcs0$ (candidate 2).

\begin{figure}
\epsfysize 3.4truein
\epsffile{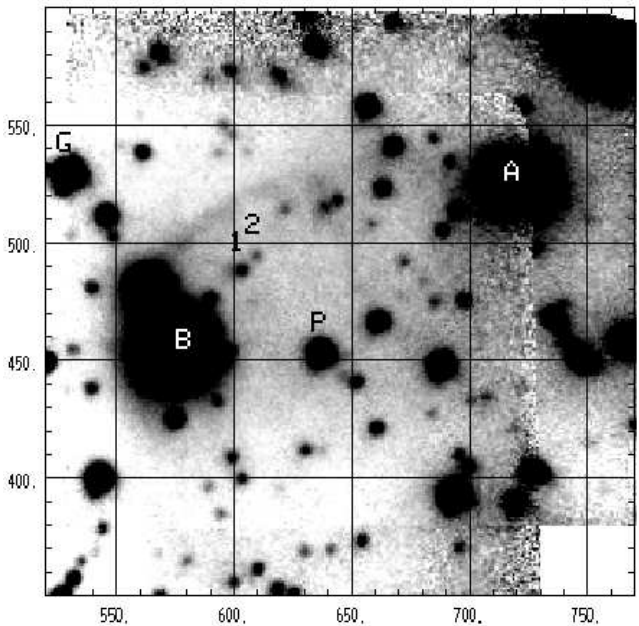}
\caption{V-band image of the core region of CTB 80.  The image is a mosaic of two 
separate observations lasting 8,173 seconds in total.  The star identifications 
refer to those listed in Blair \& Schild (1985).
(1 pixel = $0\farcs23$)}
\label{vband}
\end{figure}

\begin{figure}
\epsfysize 3.4truein
\epsffile{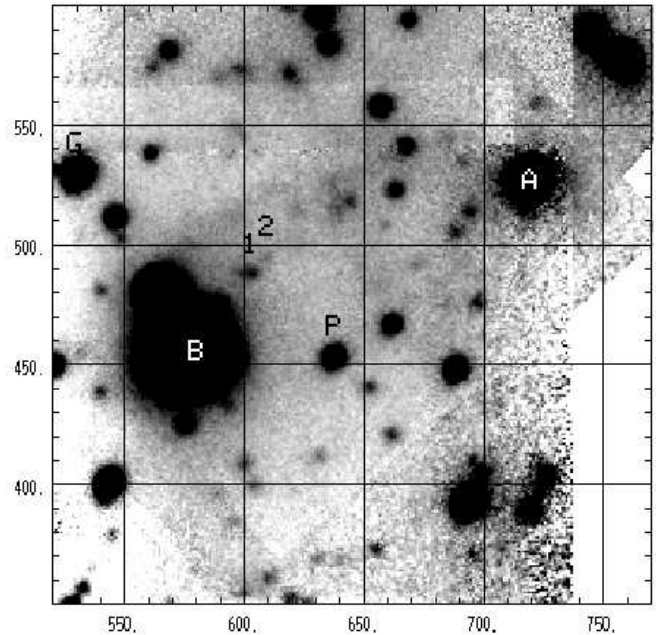}
\caption{B-band image of the core region of CTB 80. The image was made from the 
B-band observations listed in Table 1, and is equivalent to a total exposure of
16,456 seconds. The star identifications refer to those listed in 
Blair \& Schild (1985). (1 pixel = $0\farcs23$)}
\label{bband}
\end{figure}

Estimates of the distance to the pulsar vary between about 1.4 kpc (from
dispersion measure estimates, Kulkarni et al.  1988) to 3 kpc (studies of the
infrared shell, Fesen et al.  1988).  Following Safi-Harb et al.  (1995) we
adopt 2.5 kpc as a compromise between these extremes.  Reddening estimates also
vary and lie within the range E(B-V) = 0.8--1.4 (Angerhofer et al.  1980; Blair
et al.  1984).  We use a value E(B-V)= 1.0 which also gives a distance to the
pulsar of about 2.5 kpc (\cite{blair85}).  The measured values are equivalent to
a star of absolute magnitude $M_v=5.30$ and intrinsic colour
(B-V)$_{\rm{o}}=0.77$ (candidate 1) and $M_v=6.29$, (B-V)$_{\rm{o}}=0.86$
(candidate 2).

\begin{table*}
\begin{center}
\caption[]{Photometry for bright sources and pulsar candidates in the CTB 80 field.
(The values for star A are taken from Blair \& Schild (1985)).  Values for $M_v$ 
\& (B-V)$_{\rm{o}}$ are
claculated asuming a distance to the object of 2.5 kpc.}
\label{phot}
\begin{tabular}{llllllll}
\noalign{\smallskip}
\hline
\noalign{\smallskip}
        &  R.A.(1950)  & Dec. (1950) &  & & & &  \\
 Source &   $h$  $m$  $s$ &  $\degr$\  $ \arcmin$  $ \arcsec$ & V      & B & (B-V)  & $M_v$ & (B-V)$_{\rm{o}}$    \\
\noalign{\smallskip}
\hline\noalign{\smallskip}
A   & 19 51 02.10 &  32 45 15.0 & $14.95\pm0.20$   &  $16.86\pm0.20$  & $1.91$         &      &\\
1   & 19 51 02.57 &  32 44 50.42 & $20.29\pm0.30$   & $22.06\pm0.20$  & $1.77\pm0.30$  & 5.30 &0.77\\
2   & 19 51 02.47 &  32 44 51.83 & $21.28\pm0.30$   & $23.14\pm0.30$  & $1.86\pm0.30$  & 6.29 &0.86\\
4   & 19 51 02.55 &  32 44 49.25 & $22.08\pm0.20$   & $22.87\pm0.20$  & $0.79\pm0.30$  & 7.09 &-0.21\\
5   & 19 51 02.65 &  32 44 48.92 & $22.55\pm0.30$   & $23.12\pm0.20$  & $0.57\pm0.30$  & 7.56 &-0.43\\
6   & 19 51 02.48 &  32 44 47.63 & $23.05\pm0.30$   & $23.01\pm0.20$  & $-0.04\pm0.30$ & 8.06 &-1.04\\
\noalign{\smallskip}
\hline
\end{tabular}
\end{center}
\end{table*}

\begin{figure}
\epsfysize 3.4truein
\epsffile{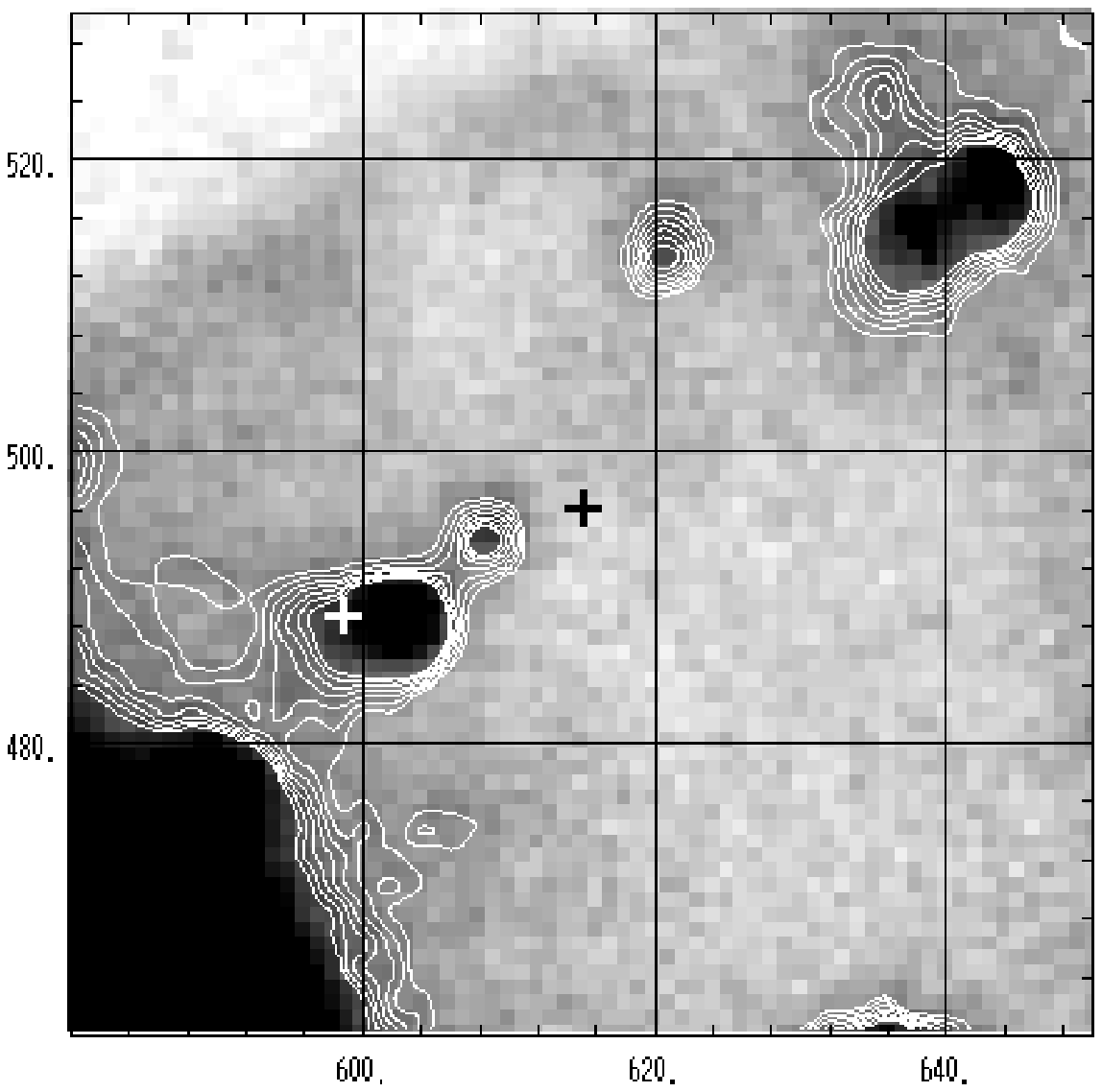}
\caption{V+B image of the central region of CTB 80. Blair and Schild's candidates
 1 and 2 can be seen toward the centre of the image.  The radio (white cross)
  and X-ray (black cross) source positions are marked.
(1 pixel = $0\farcs23$)}
\label{total}
\end{figure}

Photon times were extracted from apertures centred on the two pulsar candidates.
No pulsations were seen, to the 1\% significance level, in either the B- or
V-band data sets.  An upper limit for the pulsed fraction was calculated for the
B and V emission assuming a duty cycle in the optical of 50\%.  These upper
limits are conservative as pulse profiles in the X-ray
(\cite{safi95}) suggest a much lower duty cycle.  The 3-$\sigma$ upper limits
were found to be $m_{v,\rm{pulsed}} > 23.3\pm0.41$, $m_{b,\rm{pulsed}} > 24.2\pm0.35$,
for candidate 1, corresponding to pulsed fractions $<6\%$ in V and $<11\%$ in B.
For candidate 2 we found $m_{v,\rm{pulsed}} > 23.6\pm0.41$, $m_{b,\rm{pulsed}} > 
24.2\pm0.35$, corresponding to pulsed fractions $<12\%$ in V and $<32\%$ in B.
These limits were calculated from the longest series of data taken with a given 
filter between resets of the data-collection system (see Section 2), a single 
data file (4,417 seconds) in the case of V and the sum of three files (13,456 
seconds) in the case of B.

In Fig. 3 candidate 1 appears to be extended in the direction of the radio pulsar
position, a feature that can be seen in
both V and B data sets separately.  

An accurate PSF for the image was determined using the external IRAF package 
DAOPHOTX, and used to remove the candidate 1 and 2 stars from the image.  We
found that the extension was consistent with two point sources. Fig. 4 shows
the extension along with a another point-like feature nearby (we have labelled
these candidates $\lq$4',$\lq$5' and $\lq$6' (Fig. 4).

We have
tried to assign a magnitude and colour to each of these features, although 
their low signal-to-noise ratio and
proximity to a brighter source makes this difficult to do with any accuracy.
The photometry for these sources is included in Table 3.  Absolute magnitudes
and intrinsic colours are calculated assuming a distance of 2.5 kpc to the
pulsar.

Times were extracted from all these candidates and analysed using the $Z_{n}^{2}$
statistic and their Fourier power spectrum.  Again, no pulsations were seen to
the 1\% significance level.

Fesen \& Gull's (1985)
H$\alpha$+[NII] map of the core shows that some nebulosity is coincident with
the candidate positions, although little was detected by Blair \& Schild (1985).
Spectra of CTB 80 filaments (\cite{blair84}) show strong [OI], H$\alpha$, [NII]
and [SII] emission lines, but these are removed by our filters.  Emission lines
in the B and V bands are much weaker, although our images do show some evidence
of nebulosity. We cannot rule out the possibility that these fainter point-like
sources are emission knots in the nebula.

\begin{figure} 
\epsfysize 3truein 
\epsffile{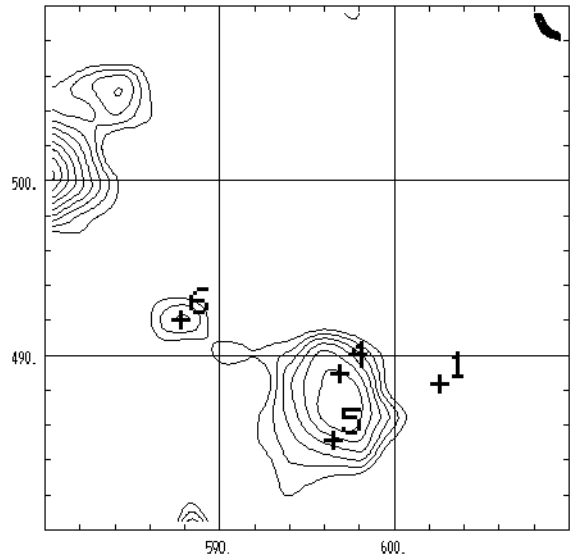}
\caption{Contour plot of the region surrounding candidate 1 in the V-band image
of Fig.1.  The position of the radio pulsar (Foster et al.  1994) is marked with
a cross.  (1 pixel = $0\farcs23$)} 
\label{vext} 
\end{figure}

Finally, we extracted photon times from apertures of radius $0\farcs81$ centred
on a two-dimensional grid of points separated by $0\farcs46$ on the 1997 data.
The grid covered an area $5\farcs5 \times 5\farcs5$ centred close to the first
candidate position.  The photon times were reduced to the solar-system
barycentre as before, and the $Z_{1}^{2}$ statistic calculated for each aperture
using the ephemeris in Table 2.  No evidence for the pulsar was found.

\section{Conclusions}
On the basis of luminosity, colour and timing we conclude that neither of the
proposed optical counterparts are the pulsar PSR B1951+32. 

We note that the slight blue extension to candidate 1 shows the expected
characteristics of a hot object.  We would not expect to be sensitive to thermal
radiation from the neutron star surface itself, as we estimate that a pulsar with T
$\approx 10^6$ K at the distance of CTB 80 would have a magnitude in the range
$m_v$=30--31.  We may, however, be able to see emission from any hot circumstellar
material.

From phenomenolgical models of pulsar optical emission (Pacini and Salvati
1983, 1987) we can derive estimated magnitudes for the emission.  For PSR
1951+32 this leads to values in the range 24--26.  However recent observations of
other middle-aged pulsars, 0656+14 and Geminga (\cite{car94}, \cite{shear98}),
have indicated a magnitude considerbly in excess of that predicted by the Pacini
and Salvati model.  It is clearly very important to establish a
positive optical identification for PSR B1951+32 in order to compare emission
from pulsars of similar ages but with a range of other parameters such as
$\dot{P}$ and magnetic field.  
These observations will be crucial if we are to distinguish between the various
models for high-energy emission. In particular, the pulse shape and
fraction can be used to differentiate the polar cap and outer gap
models.

We suggest observations with the VLT, Keck or the HST in order to definitively
identify the optical counterpart of PSR B1951+32.


\begin{acknowledgements}

Goddard Space Flight Centre is thanked for the provision of their MAMA detector.
Peter O'Kane of NUI Galway is thanked for his technical assistance.  The support
of Forbairt, the Irish research and development agency, is gratefully
acknowledged.  CO'S is supported by Forbairt under their Presidential
Post-doctoral scheme.

\end{acknowledgements}

\end{document}